\documentstyle[11pt,aaspp4]{article}
\begin{document}

\title{The Optical Gravitational Lensing Experiment.\\  Investigating the
Influence of Blending on the Cepheid Distance Scale with Cepheids in
the Large Magellanic Cloud\footnote{Based on observations obtained
with the 1.3 m Warsaw Telescope at the Las Campanas Observatory of the
Carnegie Institution of Washington}}

\author{K. Z. Stanek\altaffilmark{2}}
\affil{Harvard-Smithsonian Center for Astrophysics, 60 Garden St.,
MS20, Cambridge, MA~02138}
\affil{\tt e-mail: kstanek@cfa.harvard.edu}
\altaffiltext{2}{Hubble Fellow}
\author{A. Udalski}
\affil{Warsaw University Observatory, Al. Ujazdowskie 4,
PL-00-478 Warszawa, Poland}
\affil{\tt e-mail: udalski@astrouw.edu.pl}

\begin{abstract}

We investigate the influence of blending on the Cepheid distance
scale. Blending leads to systematically low distances to galaxies
observed with {\em HST}, and therefore to systematically high
estimates of the Hubble constant $H_0$. We select a sample of 43
long-period, large-amplitude Cepheids in the Large Magellanic Cloud,
from the catalog recently released by the Optical Gravitational
Lensing Experiment. We then model the effects of blending, as observed
by {\em HST} at large distances, by adding the $V,I$-bands
contributions from nearby bright stars. We find that the derived
distance modulus would be too short, compared to the true value, by
$\sim 0.07\;$mag at distance of $12.5\;Mpc$ and by $\sim 0.35\;$mag at
distance of $25\;Mpc$. This has direct and important implications for
the Cepheid distances to galaxies observed by the {\em HST} Key
Project on the Extragalactic Distance Scale and other teams: half of
the KP sample is likely to exhibit a blending bias greater than
$0.1\;$mag.

\end{abstract}

\section{Introduction}

As the number of extragalactic Cepheids discovered with {\em HST}
continues to increase and the value of $H_0$ is sought from distances
based on these variables (Gibson et al.~1999, Saha et al.~1999), it
becomes even more important to understand various possible systematic
errors which could affect the extragalactic distance scale. Currently,
the most important systematic is a bias in the distance to the Large
Magellanic Cloud (LMC), which provides the zero-point calibration for
the Cepheid distance scale. The LMC distance is very likely $\sim$15\%
shorter than usually assumed (e.g. Udalski et al.~1999a; Stanek et
al.~1999), but it still might be considered uncertain at the $\sim
10$\% level (e.g. Jha et al.~1999).  Another possible systematic, the
metallicity dependence of the Cepheid Period-Luminosity (PL) relation,
is also very much an open issue, with the empirical determinations
ranging from 0 to $-0.4\;mag\; dex^{-1}$ (Freedman \& Madore 1990;
Sasselov et al.~1997; Kochanek ~1997; Kennicutt et al.~1998, Udalski
et al.~1999a).

In this paper we investigate a much neglected systematic, that of the
influence of blended stellar images on the derived Cepheid distances.
Although Cepheids are very bright, $M_V\sim -4$ at a period of
$10\;days$, their images when viewed in distant galaxies are likely to
be blended with other nearby, relatively bright stars.  Recently
Mochejska et al.~(1999) showed that a significant fraction of Cepheids
discovered in M31 by the DIRECT project (e.g. Kaluzny et al.~1998;
Stanek et al.~1998) were resolved into several blended stars when
viewed on the {\em HST} images.  The average FWHM on the DIRECT
project ground-based images of M31 is about $1.5\arcsec$, or $\sim
5\;pc$, which corresponds to the {\em HST}-WFPC2 resolution of
$0.1\arcsec$ for a galaxy at a distance of $10\;Mpc$.  Any luminous
star (or several of them) in a volume of that cross section through a
galaxy could be indistinguishable from the Cepheid and would
contribute to its measured flux.

In this paper we investigate the effects of stellar blending on the
Cepheid distance scale by studying Cepheids and their close neighbors
observed in the LMC by the Optical Gravitational Lensing Experiment
(OGLE: Udalski, Kubiak \& Szyma\'nski 1997). The catalog of $\sim
1300$ LMC Cepheids has been recently released by Udalski et
al.~(1999b). As the LMC is $\sim 100-500$ times closer to us than
galaxies observed by {\em HST}, ground-based resolution of $\sim
1.0\arcsec$ allows us to probe linear scales as small as $\sim
0.25\;pc$ in that galaxy.

We describe the OGLE data used in this paper in Section 2.  In Section
3 we apply these data to simulate the blending of Cepheids at various
distances.  In Section 4 we discuss the implications of our results
for the Cepheid distance scale. In Section 5 we propose further
possible studies to learn more about the effects of blending.

\section{The Data}

The data used in this paper came from two catalogs produced by the
OGLE project. The first one, with 1333 Cepheids detected in the 4.5
square degree area of central parts of the LMC, has just been released
(Udalski et al.~1999b) and it is available from the OGLE Internet
archive at {\tt http://www.astrouw.edu.pl/\~\/ftp/ogle/}. It contains
about $3.4\times10^{5}$ $BVI$ photometric measurements for the
variables, along with their derived periods, $BVI$ photometry,
astrometry and classification.

The second catalog, that of $BVI$ photometry for many millions of LMC
stars observed by the OGLE project, will be released soon (Udalski et
al.~1999c, in preparation). Its construction will be analogous to the
SMC $BVI$ maps (Udalski et al.~1998). A typical accuracy of the
photometric zero points of the LMC photometric data is about
$0.01-0.02\;$mag for all $BVI$-bands and the catalog reaches
$V\approx21\;$mag.

As our goal was to estimate the influence of blending on the Cepheid
distance scale, we selected for further investigations only 54 longest
period ($P>10\;days$), fundamental-mode Cepheids.  We further required
their $I$-band total amplitude of variations to exceed $0.4\;$mag,
which corresponds to amplitude of $\sim 0.7\;$mag in the $V$-band.
This is to reflect the fact that in distant galaxies a typical
photometric error of $\sim 0.1\;mag$ prevents discovery of low
amplitude Cepheids. We used the $I$-band amplitude criterion because
OGLE observes mostly in the $I$-band.  The amplitude cutoff reduced
the OGLE LMC sample to 47 Cepheids. Finally, we have excluded four
highly reddened Cepheids, which left us with 43 variables.  It should
be noted that because of the CCD saturation limits in the OGLE data
the longest period Cepheid in our sample has $P=31\;days$.  This
sample will be used to investigate the effects of blending on the
Cepheid distance scale.

\section{Cepheids and Their Neighbors}

The LMC is located at about $50\;kpc$, or $\mu_{0,LMC}=18.5$, from us
(for simplicity, in the rest of this paper we use the distance scale
as adopted by the {\em HST} Key Project on the Extragalactic Distance
Scale, hereafter: KP). We define two distances for which we simulate
the blending using the LMC data: $12.5\;Mpc$ ($\mu_{0}\approx 30.5$),
somewhat shorter than the median distance of $14.4\;Mpc$ for the KP
sample (see Table~1 in Ferrarese et al.~1999), and $25\;Mpc$
($\mu_{0}\approx 32.0$), roughly corresponding to the most distant
galaxies in which Cepheids were detected with {\em HST} (Gibson et
al.~1999). The FWHM of {\em HST}-WFPC2 is $\sim 0.1\arcsec$, which at
$12.5\;Mpc$ and $25\;Mpc$ corresponds physically to area of radius
$12.5\arcsec$ and $25\arcsec$ at the LMC distance.

\begin{figure}[t]
%\plotfiddle{fig1.ps}{10.5cm}{0}{100}{100}{-205}{-370}
\plotfiddle{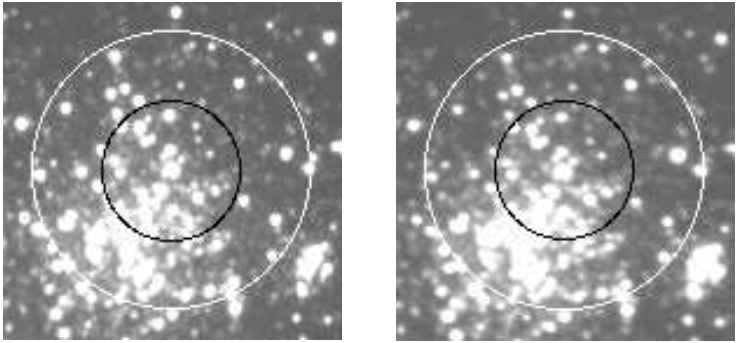}{11cm}{0}{95}{95}{-270}{-340}
\caption{Two LMC Cepheids and their neighbors shown in the $I$-band
(left panels) and the $V$-band (right panels). The images are
$1\arcmin$ in size.  Also shown are two circles corresponding to the
$0.1\arcsec$ FWHM of {\em HST}-WFPC2 camera at $12.5\;Mpc$ and
$25\;Mpc$, translating into radius of $12.5\arcsec$ and $25\arcsec$ at
the LMC distance. One pixel of the OGLE camera is $0.417\arcsec$.}
\label{fig:cepheids}
\end{figure}

In Figure~\ref{fig:cepheids} we show two OGLE LMC Cepheids and their
neighbors in the $I$-band (left panels) and the $V$-band (right
panels).  The images are $1\arcmin$ in size. Also shown are two
circles corresponding to the FWHM of {\em HST}-WFPC2 camera at
$12.5\;Mpc$ and $25\;Mpc$, i.e. $12.5\arcsec$ and $25\arcsec$ in
radius. The two Cepheids were chosen to represent two different
situations: the one shown in the top panels, LMC\_SC15 118594, has
only one bright and nearby, red neighbor at $5.4\arcsec$ from the
Cepheid, and several other, fainter neighbors further away. The second
one shown in the bottom panels, LMC\_SC11 250872, is located in a very
dense stellar region and it is probably a member of a small star
cluster. Part of the cluster light would be included in {\em
HST}-WFPC2 measurements of the Cepheid if the LMC were at $12.5\;Mpc$
while at $25\;Mpc$ almost entire cluster light would be included. As
discussed later in the paper, large amount of blended light would most
probably cause LMC\_SC11 250872 to elude detection when observed at
large distances.

The OGLE photometric catalog of stars in the LMC extends about six
magnitudes, or $\sim$0.4\% in flux (see Figure~4 of Udalski et
al.~1999b) below the bright sample of Cepheids selected in the
previous Section. We want to define a criterion to separate stars
which will contribute to the flux of a Cepheid when blended together,
from those which would contribute only to the background light in the
host galaxy. We use a lower limit of 5\% of the mean flux of the
Cepheid for a star to be included as a blend, but will discuss
different values later in the paper.

\begin{figure}[t]
\plotfiddle{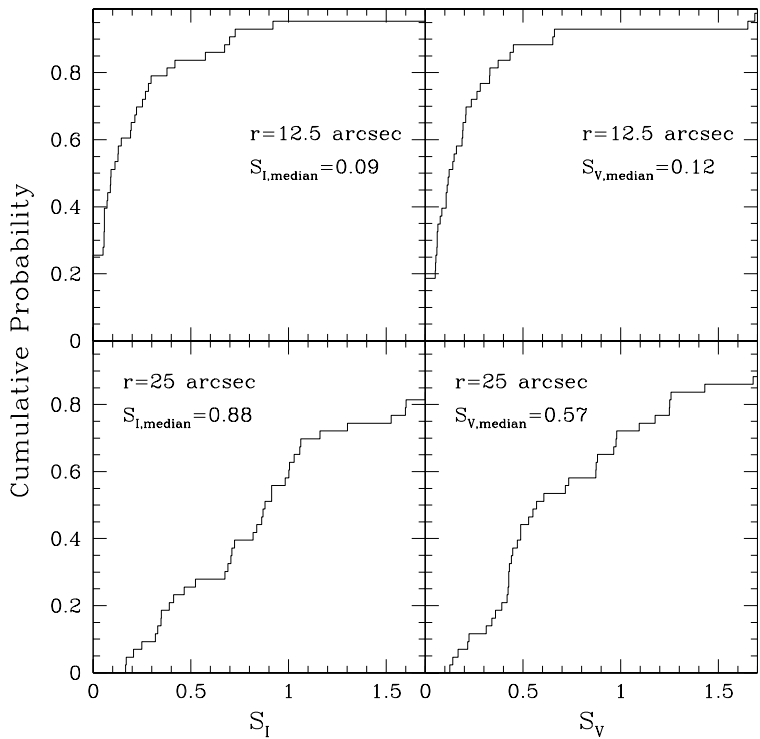}{8cm}{0}{52}{52}{-168}{-96}
\caption{The cumulative probability distribution of flux contribution
from companions $S_{I}$ (left panels) and $S_{V}$ (right panels)
within a radius of $12.5$ and $25\arcsec$ of LMC Cepheids,
corresponding to $0.1\arcsec$ FWHM of {\em HST}-WFPC2 camera at $12.5$
(upper panels) and $25\;Mpc$ (lower panels). $S_{I},S_{V}$ are defined
(Equation~1) so that when $S_{I},S_{V}=1$, the total flux from the
blended neighbors equals to the mean flux from the Cepheid.}
\label{fig:ks}
\end{figure}

We use this 5\% cutoff in evaluating the sum $S_F$ (Mochejska et
al.~1999) of all flux contributions in filter $F$ normalized to the
flux of the Cepheid:
\begin{equation}
S_F= \sum_{i=1}^{N_F}\frac{f_i}{f_C}
\label{eq:sv}
\end{equation}
where $f_i$ is the flux of the i-th companion, $f_C$ the mean
intensity flux of the OGLE LMC Cepheid and $N_F$ the total number of
companions within either $12.5\arcsec$ and $25\arcsec$ in radius.  In
Figure~\ref{fig:ks} we show the cumulative probability distribution of
flux contribution from companions $S_{I}$ (left panels) and $S_V$
(right panels) within a radius of $12.5\arcsec$ and $25\arcsec$ of LMC
Cepheids. For the smaller radius of $12.5\arcsec$ 20-25\% of our
sample is not blended and the contribution of blue blends is somewhat
stronger than that of red blends. For the larger radius of $25\arcsec$
all 43 Cepheids are blended to some extent and the contribution of red
blends is now more significant than that of blue blends. In the next
Section we attempt to use our data to quantify the effects of blending
on the Cepheid distance scale.

\section{Blending and the Cepheid Distance Scale}

We adopt the KP procedure for deriving distances, as given by
prescription in Madore \& Freedman (1991). LMC is {\em assumed} to be
at distance modulus of $\mu_{0,LMC}=18.50$, with LMC Cepheids reddened
by $E(B-V)=0.10\;$mag. When we apply this prescription to the $V,I$
data of our 43 OGLE Cepheids, we obtain values of $\mu_{0,LMC}=18.56$
and $E(B-V)=0.12\;$mag, i.e. somewhat discrepant, but basically
indicating fairly good agreement in photometry. The $rms$ scatter
around the fitted P-L relations (with their slopes fixed to that of
the KP prescription) is $0.17\;$mag in the $V$-band and $0.13\;$mag in
the $I$-band.

\begin{figure}[t]
\plotfiddle{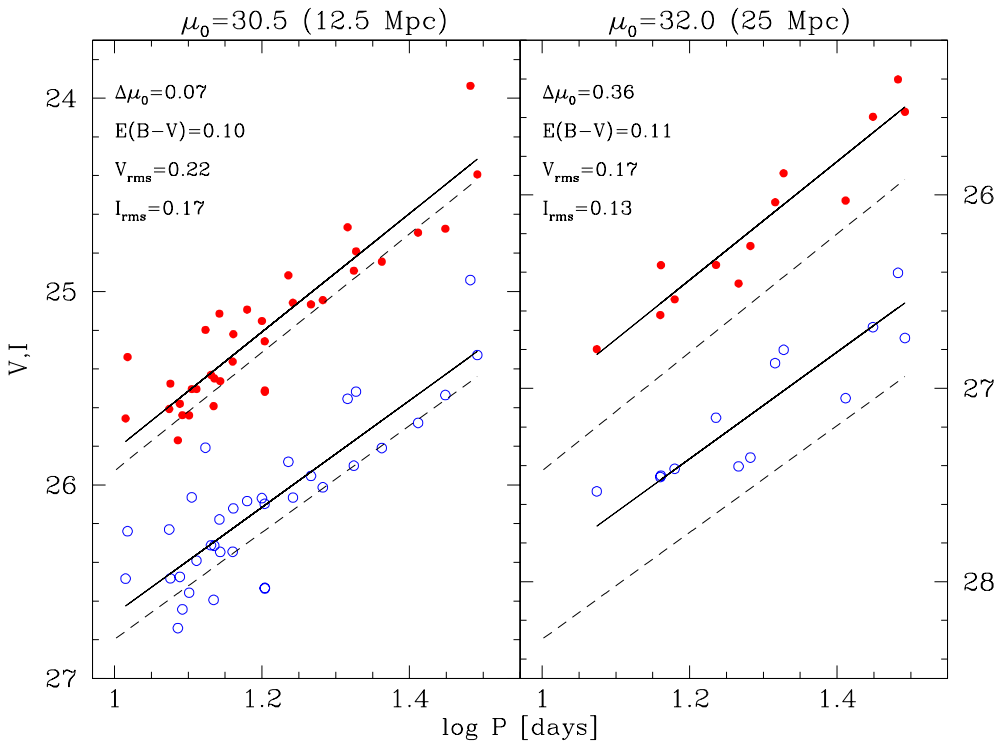}{8.2cm}{-90}{50}{50}{-207}{280}
\caption{Blending effects simulated at $12.5$ (left panels) and
$25\;Mpc$ (right panels). The continuous lines show the best-fit PL
relations when the blends are added, and the dashed lines those
without the blending. For discussion see text.}
\label{fig:pl_blend}
\end{figure}

As the next step we add contributions from the nearby stars to each
Cepheid (applying the 5\% cutoff discussed in the previous section) in
the $V$ and $I$ band separately and we repeat the distance derivation
procedure. To take into account the fact that a heavily blended
Cepheids would elude detection, we require that their blended $I$-band
total amplitude of variations exceeds $0.4\;$mag. The results for
simulated distances of $12.5$ and $25\;Mpc$ are shown in
Figure~\ref{fig:pl_blend}. This Figure shows a number of interesting
features and deserves detailed discussion.

For the simulated distance of $12.5\;Mpc$, the sample of Cepheids is
reduced to 35 and the derived distance modulus is smaller than the
``true'' (unblended) value of $\mu_{0,LMC}=18.56$ by $0.07\;$mag. The
reddening estimate is $E(B-V)=0.10\;$mag, smaller than for the
unblended sample because of the slightly larger contribution of blue
blends, as discussed in the previous Section. The $rms$ scatter around
the fitted P-L relations increases to $0.22\;$mag in the $V$-band and
$0.17\;$mag in the $I$-band. This is because while there are now
Cepheids in the sample with fairly substantial blending, it still
contains Cepheids with no blending (see Figure~\ref{fig:ks}).

The situation becomes quite dramatic at the simulated distance of
$25\;Mpc$ (right panel of Figure~\ref{fig:pl_blend}). There are only
13 Cepheids left with $I$-band amplitude larger than $0.4\;$mag, with
the shorter period (and therefore fainter) Cepheids preferentially
removed.  The derived distance modulus is smaller than the ``true''
value by $0.36\;$mag, with the reddening estimate $E(B-V)=0.11\;$mag.
What is very interesting is that the $rms$ scatter around the fitted
P-L relations now decreases to $0.17\;$mag in the $V$-band and
$0.13\;$mag in the $I$-band. This is because the sample of Cepheids,
while much smaller, is now more homogeneous when it comes to
blending. All Cepheids are to some extent blended (see
Figure~\ref{fig:ks}), with strongly blended cases removed by the high
amplitude requirement.

\begin{figure}[t]
\plotfiddle{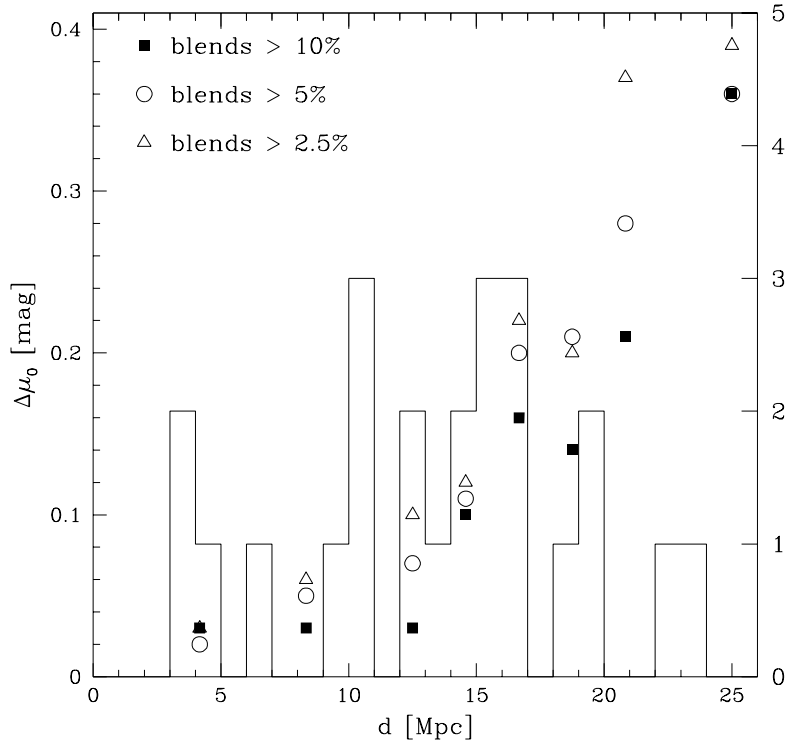}{8cm}{0}{52}{52}{-168}{-96}
\caption{Blending difference between the true and measured distances
modulus shown as a function of simulated distance. Three different
cutoffs for blending are investigated (see text for discussion): 10, 5
and 2.5\% of the mean intensity flux of the Cepheid. Also shown is the
histogram of distances for 24 galaxies for which Cepheid {\em HST}
distances were measured or re-measured by the KP (Ferrarese et
al.~1999).}
\label{fig:dist}
\end{figure}

Since there is such a dramatic difference in blending between the two
distances simulated so far, we decided to investigate the blending for
a larger number of distances. The results are shown in
Figure~\ref{fig:dist}, where blending difference between the true and
measured distance modulus is shown as a function of simulated
distance. Also shown is the histogram of distances for 24 galaxies for
which Cepheid distances with {\em HST} were measured or re-measured by
the KP (Ferrarese et al.~1999).

The 5\% cutoff which we employed to define blended stars is somewhat
arbitrary and in reality is most likely a function of data reduction
procedure, signal-to-noise in the images etc. We decided to
investigate the dependence of blending effects on the cutoff value by
using two additional cutoffs: 2.5\% and 10\%. The results are shown
with different symbols in Figure~\ref{fig:dist}. Clearly, the exact
value of the blending difference between the true and measured
distances modulus depends on the applied cutoff, but the overall trend
remains the same.

Looking at the histogram of distances in Figure~\ref{fig:dist} we can
see that half of the KP sample is likely to exhibit a blending bias
greater than $0.1\;$mag, and in some cases it can be as large as
$0.3\;$mag. Clearly blending influence on the Cepheid distance scale
can be potentially very large and cannot be neglected.

\section{Further Studies of Blending and Conclusions}

The study of Mochejska et al.~(1999) showed that a significant
fraction of Cepheids discovered in M31 by the DIRECT project were
resolved into several blended stars when viewed on the {\em HST}
images.  As we have shown in this paper, modelling of the blending
effects using Cepheids in the LMC possibly indicates a large,
$0.1-0.3\;$mag bias when deriving Cepheid distances to galaxies
observed with {\em HST}.  In addition, blending is a factor which
always contributes in only one direction, and therefore it will not
average out when a large sample of galaxies is considered. The sign of
the blending effect on the $H_0$ is opposite to that caused by the
lower LMC distance (e.g. Udalski et al.~1999a; Stanek et al.~1999).

We would like to point out that the blending of Cepheids is likely not
only to affect the studies of these stars in different galaxies, but
might also affect differential studies, such as that of Kennicutt et
al.~(1998) in the spiral galaxy M101. Their observed effect that
metal-rich (and therefore closer to the center of the galaxy) Cepheids
appear brighter and closer than metal-poor (and therefore further away
from the center) stars could be partially caused by the increased
blending closer to the center of the galaxy, although at this point we
make no attempt to estimate how much of this effect would be indeed
due to blending.

The bar of the LMC, where most of data discussed in this paper were
collected by OGLE, seems on average to have higher surface brightness
than a typical KP galaxy (Macri et al.~1999).  It would be desirable
to further establish the importance of blending for the Cepheid
distance scale using a variety of methods and data in a number of
different galaxies. Mochejska et al.~(2000, in preparation) are now
studying the {\em HST} archival images of a large sample of $\sim100$
Cepheids detected in M33 (Macri et al.~2000, in preparation) by the
DIRECT project. An approach analogous to that used in this paper will
be employed by Stanek \& Udalski (1999, in preparation) for a sample
of OGLE Cepheids in the Small Magellanic Cloud, which will have the
advantage of including Cepheids with periods of up to $P\sim 50\;days$
in a lower surface brightness system (de Vaucouleurs 1957).

Another approach, closer reproducing the procedure employed by the KP
when using Cepheids to determine distances, would be to use {\em HST}
images of relatively nearby galaxies, such as NGC3031 or NGC5253
(Ferrarese et al.~1999), by degrading them in resolution and
signal-to-noise as to represent much more distant galaxies.
Unfortunately, much of the data for the several closest galaxies have
been taken before the refurbishment of {\em HST} and are therefore of
inferior spatial resolution compared to later WFPC2 data.

All these studies can provide only an approximate answer to the
blending problem, as each individual galaxy can in principle be
different in its blending properties. One would like to find a way to
constrain or eliminate blending in each individual case.  As pointed
out by Mochejska et al.~(1999), one obvious solution to the problem of
blending would be to obtain data with better spatial resolution, such
as planned for the Next Generation Space Telescope ({\em NGST}). While
there will be desire to use {\em NGST} to observe much more distant
galaxies than with {\em HST}, it would be of great value to study some
of the not-so-distant ones as well.  Another possible approach would
be to try to circumvent the blending problem by developing and
applying a Period-Amplitude-Luminosity (PAL) relation for Cepheids
(Paczy\'nski 1999, private communication), together with image
subtraction techniques such as that developed by Alard \& Lupton
(1998).

\acknowledgments{We appreciate helpful discussions and comments on
this work by Peter Garnavich, John Huchra, Lucas Macri, Barbara
Mochejska, Bohdan Paczy\'nski and Dimitar Sasselov. Support for KZS
was provided by NASA through Hubble Fellowship grant HF-01124.01-99A
from the Space Telescope Science Institute, which is operated by the
Association of Universities for Research in Astronomy, Inc., under
NASA contract NAS5-26555.  AU acknowledges support by the Polish KBN
grant 2P03D00814. Partial support for the OGLE project was provided
with the NSF grants AST-9530478 and AST-9820314 to B.~Paczy\'nski.}

\end{document}